\documentclass[submission,Phys]{SciPost}%[submission,Phys]{SciPost}

\usepackage[english]{babel}
\usepackage{graphicx}% Include figure files
\usepackage{dcolumn}% Align table columns on decimal point
\usepackage{bm}% bold math
\usepackage{units}
\usepackage{amsmath}
\usepackage{amssymb}

\usepackage[normalem]{ulem}
\usepackage{color}
\newcounter{auxFootnote}

\begin{document}

\begin{center}{\Large \textbf{Cooling phonon modes of a Bose condensate with uniform few body losses
}}\end{center}

% TODO: write the author list here. Use initials + surname format.                                                              
% Separate subsequent authors by a comma, omit comma at the end of the list.                                                    
% Mark the corresponding author with a superscript *.                                                                           
\begin{center}
I. Bouchoule\textsuperscript{1*},
M. Schemmer\textsuperscript{1},
C. Henkel\textsuperscript{2}
\end{center}

\begin{center}
{\bf 1} Laboratoire Charles Fabry, Institut d'Optique, CNRS, Universit\'e Paris Sud 11,\\
 2 Avenue Augustin Fresnel, 91127 Palaiseau Cedex, France
\\
{\bf 2} Institute of Physics and Astronomy,
University of Potsdam, \\
Karl-Liebknecht-Str.\ 24/25,
 14476 Potsdam, Germany
\\
* isabelle.bouchoule@institutoptique.fr
\end{center}

\begin{center}
\today
\end{center}

% For convenience during refereeing: line numbers                                                                               
%\linenumbers                 

\section*{Abstract}
{\bf We present a general analysis of 
the cooling produced by losses on condensates or quasi-condensates. 
We study how  
the occupations  of the collective phonon modes evolve in time, 
assuming that the loss process is  slow enough so that
each mode adiabatically follows 
the decrease of the mean density. 
The theory is valid for any loss process whose rate is proportional
to the $j$th power of the density, but otherwise spatially uniform.
We cover both homogeneous gases and systems
confined in a smooth potential.
For a low-dimensional gas, we can take 
into account the modified equation of state 
due to the broadening of the {cloud width} along the tightly confined
directions, which occurs for large interactions.
We find that at large times, the temperature decreases
proportionally to the energy scale $mc^2$, where
$m$ is the mass of the particles and $c$ the sound velocity.
We compute the asymptotic ratio of these two quantities
for different limiting cases: a homogeneous gas in any
dimension and a one-dimensional gas in a harmonic trap.}

%% \vspace{10pt}
%% \noindent\rule{\textwidth}{1pt}
%% \tableofcontents\thispagestyle{fancy}
%% \noindent\rule{\textwidth}{1pt}
%% \vspace{10pt}

%\clearpage

%%%%%%%%%%%%%%%%%%%%%%%%%%%%%%%%%
%%% Introduction%%%%%%%%%%%%%%%%%
%%%%%%%%%%%%%%%%%%%%%%%%%%%%%%%%%

\section{Introduction}
\label{sec:intro}
Despite their extensive use as quantum simulators or for 
quantum sensing, the temperatures reached in
ultracold gases are not fully understood.
Careful analyses of the cooling mechanisms have a long
tradition in the cold atoms community, and the corresponding
temperature limits constitute important benchmarks.
The role of atom losses, however,
is not yet elucidated,
although such processes
often play a role in quantum gas experiments.
Different loss processes may occur. 
One-body processes
are always present, their origin could be for instance  
a collision with a hot atom from the residual vapour.
The familiar method of evaporative cooling 
involves losses that depend on the particle energy,
a case we exclude in this paper.
For clouds trapped in an internal state which is not the 
lowest energy state, such as low-field seekers in a 
magnetic trap, two-body (spin flip) collisions may provide 
significant loss. 
Finally, three-body processes where atoms recombine into 
strongly bound dimers are 
always present and are
often 
the dominant loss mechanism.
The effect of one-body losses
for an  ideal Bose gas was investigated in~\cite{schmidutz_quantum_2014}.
Loss processes involving more than one body are 
  a source of heating for trapped thermal clouds,
  since they remove preferentially atoms
in dense regions where the potential energy is low~\cite{weber_three-body_2003}.
Here we are interested in the effect of losses in Bose condensates or quasi-condensates,
and we focus
  on low energy collective modes, whose physics is governed by  interactions between atoms.

One-body losses have recently been investigated
for one-dimensional (1D) quasi-con\-den\-sates \cite{rauer_cooling_2016,grisins_degenerate_2016,johnson_long-lived_2017,schemmer_monte_2017}.   
Quasi-condensates characterise weakly interacting 1D Bose gases at 
low enough temperature: repulsive interactions 
prevent large density fluctuations 
such that the gas resembles locally 
a Bose Einstein condensate (BEC), although it does not sustain true 
long-range order~\cite{Petrov_2000, AlKhawaja_2003c}.
The above studies have focussed on
low-energy excitations in the gas, the 
phonon modes.
These correspond to hydrodynamic waves propagating in 
the condensate, where long-wavelength phase (or velocity) modulations are
coupled to  density modulations.
On the one hand, 
losses reduce density fluctuations and thus
remove  interaction energy from each phonon mode.
This decrease in energy, and thus of quasiparticle occupation,
amounts to a cooling of the modes.
On the other hand, the shot noise due to the discrete 
nature of losses 
feeds additional density fluctuations into the gas.
This increases the energy per mode and amounts to heating.
Theoretical studies~\cite{grisins_degenerate_2016,johnson_long-lived_2017,schemmer_monte_2017}, 
valid for one-body losses in 1D homogeneous gases, 
predict that as a net result of these competing
processes, the system is cooling down in such a way that
the ratio between temperature $k_B T$
and the chemical potential $\mu$ becomes asymptotically 
a constant (equal to 1).
Many questions remain open. For instance,
  the role of longitudinal confinement has not been elucidated.
Moreover, theoretical predictions for higher-body loss processes are 
lacking, although
cooling by three-body losses was recently demonstrated experimentally~\cite{schemmer_cooling_2018}.

In this paper, we generalise
the theoretical results  for one-body losses in homogeneous 1D gases and
extend the analysis to a
BEC or 
a quasi\-condensate in any dimension, for    
any $j$-body loss process, and for homogeneous gases as well as clouds confined in a smoothly
varying trapping potential.
We concentrate on phonon modes and
the loss rate is assumed
small enough to ensure adiabatic following of each mode.  
Low-dimensional systems are realised experimentally by freezing the transverse
  degrees of freedom with a strong transverse confinement. 
  However, in many experiments
  the interaction energy is not negligible 
  compared to the transverse excitation frequencies such that the freezing is not perfect.
  The interactions then broaden the wave function in the transverse
  directions,
  and
  longitudinal phonon modes are associated
  with transverse breathing~\cite{Stringari_1998,salasnich_effective_2002, fuchs_hydrodynamic_2003}.
Our theory can take this into account with a modified
  equation of state:
 the quantities
$\mu$ and $m c^2$, where $m$ is the atomic mass and $c$ the sound velocity, equal for a strong
transverse confinement, no longer coincide.
We find that the evolution produced by losses
is better described 
by a constant ratio $k_BT/(mc^2)$ instead of $k_B T/\mu$.
%, 
%as observed experimentally in our group~\cite{schemmer_cooling_2018}.
The asymptotic ratio $k_B T / (m c^2)$ is computed for a few 
examples.
Predictions from this paper have been tested successfully against
  recent experimental
  results obtained  at Laboratoire Charles Fabry
  on the effect of three-body losses in a harmonically confined 1D Bose
  gas~\cite{schemmer_cooling_2018}.
 %: Theoretical predictions  are in
%  good agreement with experimental data.} 
  
  %% Quantitative predictions using the equations derived in this paper
  %% are in agreement with recent experimental
  %% results obtained at Laboratoire Charles Fabry
  %% on the effect of 3-body losses in a harmonically confined 1D Bose
  %% gases~\cite{schemmer_cooling_2018}.}

\section{Model}

We consider a condensate, or quasi-condensate, in dimension $d=1,2$ or $3$.
The gas is either 
homogeneous or trapped
in a smoothly
varying  potential $V({\bf r})$.
We assume it is subject 
to a $j$-body loss process of rate constant $\kappa_j$: 
the number of atoms lost per unit time and unit volume 
is $\kappa_j n^{j}$ where $n$ is the density. 
This density
includes fluctuations of quantum and thermal nature, and
its average profile is denoted $n_0({\bf r}, t)$.
Instead of using involved powerful theoretical techniques such as
the truncated Wigner approach~\cite{norrie_three-body_2006,drummond_functional_2013},
we compute the effect of losses
in this paper with a spatially coarse-grained approach
that does not rely on involved theory
and in which the approximations are made transparent.
For the same pedagogical reason, we explicitly construct the 
phase-density
representation of the collective excitations of the gas,
in a similar way as is done for instance in~\cite{mora_extension_2003}.

\subsection{Stochastic dynamics of the particle density}
\label{s:dN-per-pixel}

Let us first consider the sole effect of losses and fix
a cell of the gas of volume $\Delta$, 
small enough so that the density of the (quasi\-)condensate is about 
homogeneous in this volume, but large enough to accommodate many atoms.
The atom number in the cell is $N = N_0 + \delta N$ 
where $N_0 = n_0\Delta$ and $\delta N\ll N_0$  
since the gas lies in the {(quasi\-)}condensate regime. (We drop the
position dependence $n_0 = n_0( {\bf r} )$ for the moment.)
Since typical values of $\delta N$ are much smaller than $N_0$,
one can assume without consequence
that $\delta N$ is a variable that takes discrete values between $-\infty$ and $\infty$. 
Hence, one can define
a phase operator $\theta$, whose eigenvalues span the interval
$[0,2\pi[$ and
that is canonically
conjugate to $\delta N$. 
Losses will affect both the density fluctuations and the phase fluctuations.

We  first concentrate on the effect of losses on density fluctuations.
Consider a time step $dt$, 
small enough that the change $dN$ in atom number
is much smaller than $N$, but large enough such that $dN$ is much larger than $1$. 
After the time step, we have
\begin{equation}
dN = - K_j N^{j} dt + d\xi
\label{eq:stoch-dN}
\end{equation}
where $K_j = \kappa_j / \Delta^{j-1}$. Here, 
$d\xi$ is a random number with vanishing mean value
that translates the shot noise associated with the statistical nature of 
losses.
%~\footnote{Within the approximations
%made in the following, the Ito and  Stratonovich formalisms
%are equivalent.}.
The number of loss events during the small step $dt$
is Poisson distributed so that the variance of  $d\xi$
relates to the mean number of lost atoms by
\begin{equation}
\langle d\xi^2\rangle = 
j K_j N^{j} dt \simeq j K_j N_0^{j} dt
\,,
\label{eq:dxi2}
\end{equation}
the factor $j$
coming from the fact that at each event, $j$ atoms are lost.
The evolution of fluctuations in the atom number is obtained from 
$d\delta N = dN - dN_0$,
where $dN_0$ is the change of the mean number,
equal to $dN_0 = -K_j N_0^{j} dt$ in 
the lowest order in $\delta N$.
Expanding $N^{j}$ in Eq.(\ref{eq:stoch-dN})
to first order in $\delta N$,
we obtain the following evolution for the density fluctuation
$\delta n = \delta N/\Delta$:
\begin{equation}
  d\delta n = -j \kappa_j n_0^{j-1}\delta n\, dt  +d\eta
  \label{eq.ddeltan}
\end{equation}
where  $d\eta=d\xi/\Delta$ is a random  variable of  variance 
$\langle d\eta^2\rangle = j \kappa_j n_0^{j} dt / \Delta$. 
The first term in the r.h.s, the drift term, decreases
the density fluctuations. It will thus reduce the interaction energy 
associated to
fluctuations in the gas and produce cooling.
The second term on the other hand increases the
density fluctuations in the gas which leads to heating.

\subsection{Shot noise and phase broadening}

We now compute the effect of losses on the phase fluctuations, following
an approach similar to
Ref.\cite{korotkov_continuous_1999}.
For this purpose, 
one imagines that
one records the number of lost 
atoms during $dt$. 
This measurement increases the knowledge 
about $N$, and thus $\delta N$.
To quantify this increase of knowledge, we use the Bayes formula 
\begin{equation}
  {\rm P}(\delta N|N_l)=\frac{{\rm P}(\delta N)}{\int d(\delta N') {\rm P}(N_l|\delta N')} {\rm P}(N_l|\delta N),
  \label{eq.bayes}
\end{equation}
where ${\rm P}(\delta N)$ is  the initial probability of having an atom number
$N = N_0 + \delta N$, and
${\rm P}(N_l|\delta N)$ is the probability that a number $N_l$
of atoms will be lost, given that the initial atom number was 
$N_0 + \delta N$.
Finally, ${\rm P}(\delta N|N_l)$ is the probability that the 
final
number 
is $N_0 -N_l+ \delta N$,
knowing the fact that $N_l$ atom have been lost.
As argued above, the Poissonian nature of the loss process 
and the assumption that the number of
lost atoms is large compare to one, imply the Gaussian
distribution
\begin{equation}
{\rm P}(N_l|\delta N) 
\simeq 
\frac{ 1 }{ \sqrt{2\pi}\sigma_l }
e^{-(N_l - K_j N^{j} dt)^2 / (2\sigma_l^2)}
\,,
\label{eq:Pnl}
\end{equation}
where
$N=N_0+\delta N$ and $\sigma_l^2 = j K_j N_0^{j} dt$.
Expanding $N^{j}$ around $N_0^{j}$ and introducing 
$\overline{\delta N} = N_l/(j K_j N_0^{j-1} dt) - N_0/j$,
one has 
\begin{eqnarray}
\frac{ (N_l - K_j N^{j} dt)^2 }{ \sigma_l^2 } &\simeq&
\frac{ (\overline{\delta N} - \delta N)^2 }{ \sigma_{\delta N}^2 }
%\label{eq:}
\end{eqnarray}
where
\begin{equation}
  \sigma_{\delta N}^2 = \frac{ N_0 }{ j K_j N_0^{j-1} dt }
  \,.
\end{equation}
Thus, according to Eq.(\ref{eq.bayes}), the width of the distribution 
in $\delta N$ 
is multiplied by a function of rms width $\sigma_{\delta N}$ 
after recording the number of lost atoms. 
This narrows the number distribution and must be associated with a
broadening in the conjugate variable, $\theta$, lest the
uncertainty relations are violated. The phase broadening must
be equal to 
\begin{equation}
  \langle d\theta^2\rangle  =
  \frac{ 1 }{ 4 \sigma_{\delta N}^2 }
  =   
  \frac{ j \kappa_j n_0^{j-1} }{ 4 n_0 \Delta }
  dt
  \,.
\label{eq.phasespread}
\end{equation}
This spreading of the phase results from  the shot noise
in the loss process. 

In the following, keeping in mind that only length scales 
larger than the interparticle distance 
have to be considered, we go to the continuous limit. The factors
$1/\Delta$ in the variance for $d\eta$ in Eq.(\ref{eq.ddeltan})
and in the phase diffusion of Eq.(\ref{eq.phasespread})
then turn into
\begin{eqnarray}
\langle d\eta({\bf r})d\eta({\bf r'})\rangle &=&
j \kappa_j n_0^{j} \delta({\bf r-r'}) dt
\\
\langle d\theta({\bf r}) d\theta({\bf r'}) \rangle &=& 
\frac{ j }{ 4 } \kappa_j n_0^{j-2}\delta({\bf r-r'}) dt
\label{eq:density-phase-noise}
\end{eqnarray}
Both diffusion terms are due to
the quantised nature of the bosonic field, namely the discreteness of 
atoms. Their effects become negligible compared to the drift term 
in Eq.~(\ref{eq.ddeltan}) in the classical field limit, i.e.
$n_0\rightarrow \infty$ at fixed typical density
fluctuations $\delta n/n_0$.
Note finally that these results could also have been obtained using a 
truncated Wigner approach~
\cite{drummond_functional_2013,norrie_three-body_2006}, using approximations based on the
relation $\delta n\ll n_0$.

Before going on, let us make a remark
concerning gases in reduced dimension. An effective  1D (resp.\ 2D)
gas is obtained using a strong transverse confinement in order to freeze the transverse degree
of freedom: the atoms are in the transverse ground state of the confining potential, of
wave function $\psi( x_\perp )$.
In the case of $j$-body losses with $j>1$, the loss process a priori modifies the transverse
shape of the cloud since it occurs preferentially at the center, where the density is the highest.
In other words, it 
introduces couplings towards  transverse excitations.
We assume here the loss rate
to be much smaller than the frequency gap $\omega_\perp$ 
between the transverse ground
and first excited states. Then the coupling to transverse excitations 
has negligible effects,
and the above analysis of the effect of losses also holds for the effective 1D (resp.\ 2D) gas, provided
$\kappa_j = \kappa_j^{3D}\!\int\!d^2x_\perp 
|\psi(x_\perp)|^{2j}$ 
(resp.
$\kappa_j=\kappa_j^{3D}\!\int\! dx_\perp |\psi(x_\perp)|^{2j}$), 
where  
$\kappa_j^{3D}$ is the rate constant coefficient for the 3D gas.

\subsection{Collective excitations}
\label{s:hydrodynamic-modes}

Let us now take into account the dynamics of the gas. Under the effect 
of  losses  
the profile $n_0({\bf r}, t)$ evolves in time and, 
except for a homogeneous system,
a mean velocity field appears,
generated by a spatially dependent phase $\theta_0({\bf r},t)$.
Here we assume the loss rate is small enough so that, at any time, 
$n_0({\bf r})$ is close to the equilibrium profile. We moreover
assume the potential varies sufficiently smoothly such that
the equilibrium profile  is obtained with the local density approximation.
Then, at any time, $n_0({\bf r})$ fulfills
\begin{equation}
\mu(n_0({\bf r}))=\mu_p - V({\bf r})
\end{equation}
where $\mu(n)$ is the chemical potential of a homogeneous gas of density
$n$ and $\mu_p$ is the peak chemical potential, which fixes the 
total atom number~\footnote{The peak density is reached at the
  position ${\bf r}_p$ where   $V$ reaches its minimum value. 
We impose $V({\bf r}_p)=0$.}.
In most cases $\mu=g n$ where $g$ is the coupling constant. 
In 3D condensates, $g=4\pi\hbar^2a/m$ where
$a$ is the scattering length describing low-energy collisions.
In situations where two (resp. one) degrees of freedom are strongly confined {by a
transverse potential of frequency $\omega_\perp$},
$\mu$  depends on $a$, on the linear (resp. surface) density $n$,
and on $\omega_\perp$.
As long as $\hbar\omega_\perp \gg \mu$, the transverse cloud shape is close to that of the transverse
ground state  \footnote{We assume here that the transverse width of the cloud
fulfills $l_\perp\gg a$  such that the effect of interactions 
is well captured treating the gas as a 3D gas.}, and
one recovers the expression  $\mu=gn$ where the effective
   1D (resp. 2D) coupling constant $g$ depends only on $a$ and on $\omega_\perp$~\cite{petrov_bose-einstein_2000,olshanii_atomic_1998}.
At large densities, $\hbar \omega_\perp \sim \mu$, the transverse degrees of freedom are no longer
completely frozen: interactions {broaden} the transverse wave function,
and $\mu$ is no longer linear in $n$~\cite{salasnich_effective_2002, fuchs_hydrodynamic_2003}.
We discuss one example in Sec.\ref{s:homogeneous}.

To treat the dynamics around the average density $n_0( {\bf r}, t )$, 
a Bogoliubov approximation
is valid since the gas is in the (quasi\-)condensate regime: 
one can linearise the equations of motion in the density and phase 
fluctuations $\delta n({\bf r})$
and  $\varphi({\bf r})=\theta-\theta_0$~\cite{PitaevskiiStringari, mora_extension_2003}.
These equations involve the mean velocity field $\hbar \nabla \theta_0/m$.
Here we assume the loss rate is small enough so that
such terms are negligible.
We moreover consider only length scales much larger than the healing length.
Then, as detailed in Appendix~\ref{a:lowD-hdyn}, the dynamics of $\delta n({\bf r})$ and  $\varphi({\bf r})$ 
is governed by the 
hydrodynamic Hamiltonian
\begin{equation}
  H_{\rm{hdyn}} = \frac{\hbar^2}{2m}\int d^{d}{\bf r}\, n_0 \left( \nabla \varphi \right)^2
  +\frac{m}{2}\int d^{d}{\bf r} \frac{ c^2 }{ n_0 } \delta n^2.
  \label{eq.Hydro}
\end{equation}
Here the speed of sound $c = c( {\bf r} )$ is related 
to the local compressibility,
$m c^2 = n_0 \partial_n \mu$, 
evaluated at $n_0({\bf r})$.
At a given time,
$H_{\rm{hdyn}}$ can be recast as a collection of independent collective modes.
The collective modes are  described by the 
eigenfrequencies $\omega_\nu$ and the 
real functions $g_\nu$ [details in Appendix~\ref{SM}]. 
They obey
\begin{equation}
\nabla \cdot \big( n_0 \nabla (\frac{ c^2 }{ n_0 } g_\nu) \big)
= - \omega_\nu^2g_\nu,
	\label{eq:wave-eqn-gnu}
\end{equation}
and are normalised according to
\begin{equation}
\delta_{\nu,\nu'} = \frac{ m }{ \hbar \omega_\nu } 
\int d^{d}{\bf r} \frac{ c^2 }{ n_0 } 
g_\nu({\bf r}) g_{\nu'}({\bf r})
\,.
\label{eq:g-nu-normalisation}
\end{equation}
Then $H_{\rm{hdyn}}=\sum_\nu H_\nu$ where
 \begin{equation}
 H_\nu = \frac{ \hbar\omega_\nu }{ 2 } ({x_\nu^2} + {p_\nu^2}).
\end{equation}
The dimensionless canonically conjugate quadratures $x_\nu$ and $p_\nu$ are related to 
$\delta n$ and $\varphi$ respectively.
More precisely,
\begin{equation}
  \left \{\begin{array}{l}
  \delta n({\bf r}) = \sum_\nu x_\nu g_\nu({\bf r})
  \\[1ex]
  \displaystyle
  \varphi({\bf r}) = \frac{ m c^2 }{ n_0 }
      \sum_\nu p_\nu \frac{ g_\nu({\bf r}) }{ \hbar\omega_\nu }
  \end{array}
  \right .
  \label{deltanvsxnu}
  \end{equation}
which inverts into
\begin{equation}
    \left\{
    \begin{array}{l}
    \displaystyle
    x_\nu = \frac{m}{\hbar\omega_\nu}\int d^{d}{\bf r} 
    \frac{ c^2 }{ n_0 } \delta n({\bf r}) g_\nu({\bf r})
    \\[2ex]
    p_\nu = \int d^d{\bf r}\, \varphi({\bf r}) g_\nu({\bf r})
    \end{array}
    \right.
\label{eq.xnu}
\end{equation}
At thermal equilibrium, the energy in the mode $\nu$ is equally shared between
both quadratures and, for temperatures $T\gg \hbar \omega_\nu$, one has 
$\langle H_\nu \rangle =T$.

\section{Cooling dynamics}

\subsection{Evolution of the excitations}

Let us consider the effect of losses on the  collective modes.
The loss process modifies in time the mean density profile and thus 
the two functions of ${\bf r}$, $n_0$ and  $c$,
 that enter into the Hamiltonian Eq.~(\ref{eq.Hydro}).
 We however assume  the loss rate is very low compared to the 
mode frequency and their differences $\omega_\nu - \omega_{\nu'}$,
so that the system follows adiabatically the 
effect of these modifications. As a consequence, 
equipartition of the energy holds at all times 
for any collective mode $\nu$,
and the adiabatic invariant 
$A_\nu=\langle H_\nu\rangle /(\hbar \omega_\nu)$ 
is unaffected by the slow evolution of $n_0$. 
The dynamics of $A_\nu$ is then only due to the modifications
of $\delta n( {\bf r} )$ and $\varphi( {\bf r} )$ 
induced by the loss process (subscript $l$), namely
\begin{equation}
\frac{ d A_\nu }{ dt } = 
\frac{1}{2} \Big( 
\frac{ d\langle x_\nu^2\rangle_l }{ dt }
+ \frac{ d\langle p_\nu^2\rangle_l }{ dt }
\Big)
\label{eq.dEtilde}
\end{equation}
Injecting Eq.~(\ref{eq.ddeltan}) into
Eq.~(\ref{eq.xnu}), we obtain
for the `density quadrature'
\begin{equation}%array}
  (d x_\nu)_l = \frac{m}{\hbar\omega_\nu}\int d^d{\bf r} \,
  \frac{ c^2 }{ n_0 } g_\nu({\bf r})
 \left(
  -j \kappa_j n_0^{j-1}\delta n({\bf r}) dt + d\eta({\bf r})
  \right).
 % \nonumber
  \label{eq.dxnul}
\end{equation}%narray}
Using the mode expansion~(\ref{deltanvsxnu}) for $\delta n( {\bf r} )$ 
in the first term, we observe the appearance of
couplings between modes. In the adiabatic limit (loss rate small 
compared to mode spacing), the effect of these couplings 
is however negligible. Then, Eq.~(\ref{eq.dxnul}) leads to
\begin{equation}
  \frac{d \langle x_\nu^2 \rangle_l}{dt} =
  - \frac{ 2j \kappa_j m }{ \hbar\omega_\nu }
  \langle x_\nu^2 \rangle 
  \int\!d^d{\bf r}\, c^2 n_0^{j-2} g_\nu^2
%  \nonumber
%  \\
  + \frac{ j \kappa_j m^2 }{ (\hbar \omega_\nu)^2 }
 \int\!d^d{\bf r} \, c^4 n_0^{j-2} g_\nu^2
 \,.
\label{eq.dxnu2l}
\end{equation}%narray}
Let us now turn to the phase diffusion associated with losses. It modifies 
the width of the conjugate quadrature $p_\nu$, according to
\begin{equation}
\frac{d \langle p_\nu^2 \rangle_l}{dt} 
= 
 \frac{ j \kappa_j }{ 4 } 
\int\!d^d{\bf r} \, n_0^{j-2}  g_\nu^2
\,.
\label{eq.dpnu}
\end{equation}
The hydrodynamic modes are characterised by low energies, 
$\hbar\omega_\nu \ll m c^2$, when the speed of sound is 
evaluated in the bulk of the {(quasi\-)}condensate.
Then $d \langle p_\nu^2 \rangle_l/dt $ gives a 
contribution that scales with the small factor
$(\hbar\omega_\nu / m c^2)^2$
compared to the second term of Eq.~(\ref{eq.dxnu2l}). 
In other words one expects that the phase diffusion associated 
to the loss process gives a negligible contribution 
to the evolution of $A_\nu$
[Eq.(\ref{eq.dEtilde})] for phonon modes
\footnote{At the border of the {(quasi\-)}condensate, 
where the density becomes small,
the condition $\hbar \omega_\nu \ll m c^2$ breaks down, however. 
The effect of phase diffusion is more carefully evaluated 
in Sec.\ref{s:harmonic}.}.

We see from Eq.(\ref{eq.dxnu2l}) that the adiabatic invariant 
$A_\nu$ is actually changed by 
$j$-body losses.
We now show that the decrease in the energy per mode
$\langle H_\nu\rangle$ is better captured by the energy scale 
associated with the speed of sound, as their ratio will 
{converge towards}
a constant during the loss process.
More precisely, we introduce
\begin{equation}
y_\nu = \frac{ \langle H_\nu\rangle }{ m c_p^2 }
\simeq \frac{ k_B T_\nu }{ m c_p^2 }
%\label{eq:}
\end{equation}
where $c_p$ is the speed of sound evaluated at the peak density
$n_p$. 
The second expression is valid
as long as the phonon modes stay in the classical regime,
$\langle H_\nu\rangle\gg \hbar \omega_\nu$.
From Eq.~(\ref{eq.dEtilde}) and (\ref{eq.dxnu2l}), neglecting the 
contribution of~Eq.(\ref{eq.dpnu}),  we immediately obtain   
\begin{equation}
\frac{d}{dt} y_\nu = 
\kappa_j n_p^{j-1}
\left[ 
	- (j{\cal A} - {\cal C}) y_\nu 
	+ j{\cal B} 
\right]
\label{eq.ydot}
\end{equation}     
where the dimensionless parameters ${\cal A},{\cal B}$ and ${\cal C}$ are 
\begin{eqnarray}
  {\cal A} &=& \frac{m}{\hbar \omega_\nu}
  \int\!d^{d}{\bf r} 
  \frac{ c^2 n_0^{j-2} }{ n_p^{j-1} } g_\nu^2({\bf r})
  \,, 
  \label{eq.calA}
\\
  {\cal B} &=& \frac{m}{2\hbar\omega_\nu}
  \int\!d^{d}{\bf r} \frac{ c^4 n_0^{j-2} }{ c_p^2 n_p^{j-1} } g_\nu^2({\bf r})
  \,, 
  \label{eq.calB}
\\
  {\cal C} &=&
  \frac{ d\ln (m c_p^2 / \hbar\omega_\nu ) }{ dN_{\rm tot} }
  \int\!d^d{\bf r} \frac{ n_0^{j} }{ n_p^{j-1} }
  \,.
  \label{eq.calC}
\end{eqnarray}
In general, all of them depend on $\nu$ but we omit the index $\nu$ for compactness.
The term ${\cal A}$ is the rate of decrease of $y_\nu$ induced by the reduction
of the density fluctuations under the loss process, normalised to
$\kappa_j n_p^{j-1}$. The term ${\cal B}$ originates from the
additional density fluctuations induced by the
stochastic nature of the losses. The term ${\cal C}$ arises from
the time dependence of the ratio
$m c_p^2 / \hbar\omega_\nu$. It is computed  using the dependence of
$m c_p^2 / \hbar\omega_\nu$ on the total atom number, the latter  evolving according to
\begin{equation}
\frac{ dN_{\rm tot} }{ dt } = - \int\!d^d{\bf r}\, \kappa_j n_0^{j}
\,.
\end{equation}
Eqs.~(\ref{eq.ydot}--\ref{eq.calC}) constitute the main results of this paper.
They have been solved numerically for the experimental parameters corresponding to
  the data of~\cite{schemmer_cooling_2018} ($j=3$ and anisotropic harmonic confinement)
  and their predictions compare very well with experimental results.

We would like at this stage to make a few comments about these equations.
First, the factor $\hbar$, although it appears explicitly in the equations, is not relevant
since it is canceled by the $\hbar$ contained in the normalisation~(\ref{eq:g-nu-normalisation}) of the
mode functions $g_\nu$.
Second, we note that  ${\cal A}$, ${\cal B}$ and ${\cal C}$ are intensive
parameters: they are  invariant by a scaling 
transformation $V({\bf r})\rightarrow V(\lambda {\bf r})$ and depend
only on the peak density $n_p$ and on the shape of the potential.
Finally,  Eqs.~(\ref{eq.ydot}--\ref{eq.calC}) depend on $\nu$ and 
it is possible that  the lossy (quasi-)condensate evolves into
a non-thermal state where different modes acquire different temperatures.
Such a non-thermal state of the gas is permitted within the linearised approach
where modes are decoupled. In the examples studied below, however,
it turns out that all hydrodynamic modes
share about the same temperature\footnote{\label{footnote:nonthermal}\hypertarget{footnotetarget:nonthermal}{
In} the case of one-body
losses, theories that go beyond the hydrodynamic approximation 
predict non-thermal states to appear, where the
high-frequency modes reach higher temperatures than the phonon
modes~\cite{grisins_degenerate_2016,johnson_long-lived_2017}.}. \setcounter{auxFootnote}{\value{footnote}}
In the following, we investigate the consequences of Eq.~(\ref{eq.ydot}-\ref{eq.calC}), considering
different situations.

\subsection{Example: homogeneous gas}
\label{s:homogeneous}

In this case,
density $n_0$ and speed of sound $c$ are spatially
constant. 
The collective modes are sinusoidal functions, labelled by $\nu$ and of wave
  vector ${\bf k_\nu}$~\footnote{For 1D gases, $\nu=(p,\sigma)$ where $p$ is a positive integer 
    and $\sigma=c$ or $s$ depending wether we consider cosine or sine modes.
    The wave-vector is $k_\nu=2p\pi/L$ where
    $L$ is the length of the box, assuming periodic boundary conditions.
    This generalises to higher dimensions with $\nu=(p_1,\sigma_1,p_2,\sigma_2,p_3,\sigma_3)$ in 3D for instance.}.
The frequencies are given by the acoustic 
dispersion relation
$\omega_\nu = c |{\bf k_\nu}|$ and
the mode functions $g_{\nu c,s}({\bf r})$ 
are normalised to 
\begin{equation}
\int d^d{\bf r}\, g_{\nu}^2({\bf r}) = 
\frac{ \hbar \omega_\nu }{ m c^2 } n_0
\end{equation}
Then Eqs.(\ref{eq.ydot}-\ref{eq.calC}) reduce to
\begin{equation}
\frac{d}{dt} y = \kappa_j n_0^{j-1}
\left[ - y \left(
	j - \frac{ \partial \log c }{ \partial \log n_0 }
	\right) + j/2 \right]
\label{eq.dydthomo}
\end{equation}
which is the same for all modes $\nu$.
Let us consider the limit  $\mu=gn_0$, valid in 3D gases, or in low-dimensional gases with strong transverse confinement
(negligible broadening of the transverse wave function).
Then $c \propto n_0^{1/2}$ and
Eq.~(\ref{eq.dydthomo}) shows that
$y$ tends at long times towards the asymptotic value
\begin{equation}
y_{\infty} = \frac{1}{2-1/j}
\,,
\end{equation}
independent of the mode energy\hyperlink{footnotetarget:nonthermal}{\footnotemark[\value{auxFootnote}]}.
For one-body losses, one recovers the result $y_{\infty}=1$~\cite{grisins_degenerate_2016,johnson_long-lived_2017}.
In the case of 3-body losses, one finds $y_{\infty} = 3/5$.

\begin{figure}[h!]
  \centerline{\includegraphics*[width = 0.7\columnwidth]{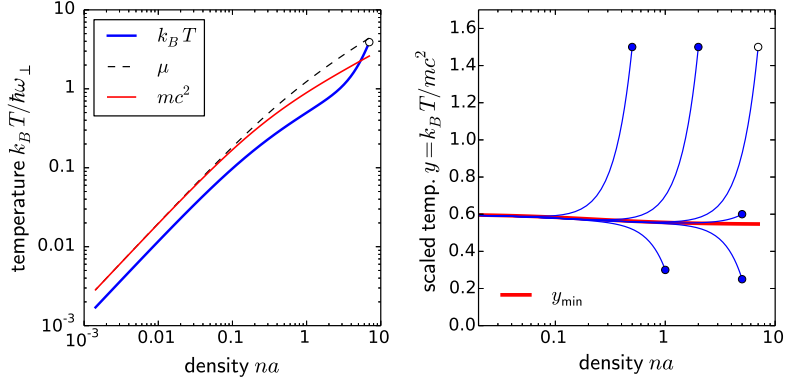}}
  \caption{
    Cooling a quasi-1D gas, homogeneous along the axial direction, by three-body losses. The
    density is initially so high that 
    transverse broadening is relevant [the chemical 
      potential 
      does not fulfill $\mu \ll \hbar\omega_\perp$].
    Left: time evolution of the temperature (thick blue), shown versus the
      time decreasing density.  Black dashed and thin red lines show the 
    intrinsic energy scales {$\mu$ and $mc^2$}.
    The system rapidly evolves into a dynamical state 
    where the temperature
    follows the energy scale $m c^2$, rather than the chemical
    potential.
    Right: evolution of the ratio $y = k_B T / mc^2$ vs.\ the 
    density. The curves correspond to different initial values
    (marked with dots, the white dot corresponding to the parameters
    on the left). 
    The thick red line shows the function $y_{\rm min}$ that gives the {positions of 
    lowest values taken by $y$} in the course of cooling. In this system
    (homogeneous along the axial direction), all hydrodynamic modes evolve
    with the same temperature. 
    }
  \label{fig.effectbroadening}
  \end{figure}

Let us now consider a quasi-low-dimensional gas, where 
transverse broadening of the wave function cannot be neglected.
The logarithmic derivative in Eq.(\ref{eq.dydthomo}) is then
no longer constant.
We will focus on the case of a quasi-1D gas, as realised experimentally for instance
in~\cite{schemmer_cooling_2018}.
The effect of the
transverse broadening is well captured by the
heuristic equation of state~\cite{salasnich_effective_2002, fuchs_hydrodynamic_2003}
\begin{equation}
\mu = \hbar\omega_\perp\left( \sqrt{1+4 n_0 a}-1 \right),  
	\label{eq:eff-mu-swelling}
\end{equation}
where $\omega_\perp$ is the frequency of the transverse confinement
and $a$ the 3D scattering length.
Inserting into  Eq.~(\ref{eq.dydthomo}), one can compute the evolution
of $y$. The transverse broadening also modifies the
rate coefficient $\kappa_j$, making it density-dependent. 
However, re-scaling the time according to
$u=\int_0^t \kappa_j(\tau)n_p^{j-1}d\tau=\ln(n_0(0)/n_0(t))$,
Eq.~(\ref{eq.dydthomo}) transforms into
\begin{equation}
{\frac{dy}{du} = 
- y \left( j - 1/2 + \frac{ n_0(0) a \, e^{-u} }{ 1 + 4 n_0(0) a \, e^{-u}} \right) + j/2}
\end{equation}
and no longer depends on $\kappa_j$.
Fig.\ref{fig.effectbroadening} shows the solution of
this differential equation in the case of 3-body losses, and
for a few initial situations, namely different values of $y$ and
$n_0 a$ (right plot). 
The asymptotic value $y=y_\infty$ is always reached at long times 
since the transverse broadening then becomes negligible.
Note that in distinction to pure 1D gases, 
the effect of transverse broadening allows the system to reach transiently
  lower {scaled} temperatures $y < y_\infty $,
even when starting at values of $y$ larger than $y_\infty$.
More precisely, let us denote 
$y_{\rm min}(n_0) = j/2/(j-1/2+an_0/(1+4an_0))$.
When starting with $y>y_{\rm min}$, the lowest value of $y$ 
is reached for some (non-vanishing) density, {and} it 
falls on the curve $y_{\rm min}$.
For $j=3$, one find that $y_{\rm min}$ varies between $y_\infty=0.6$ and $6/11\simeq 0.55$.
Thus, the coldest temperatures in the course of the loss process
never deviate by more than 10\% from the asymptotic value 0.6: the impact of transverse
swelling is relatively small.
Note that, if one considered the scaled temperature $T/\mu$ rather than $y$, much larger deviations
would appear.

\subsection{Example: 1D harmonic trap}
\label{s:harmonic}

\newcommand{\RTF}{R}

\begin{figure}[h!]
  \centerline{\includegraphics[width=0.65\columnwidth]{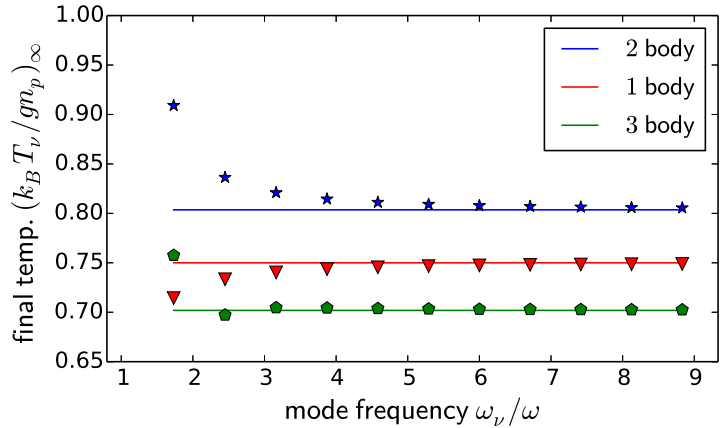}}
  \caption{
    Asymptotic ratio $y_\infty=k_BT/mc^2$ for hydrodynamic collective modes
    of a 1D quasi-condensate confined in 
    a harmonic trap, for 1-body (red), 2-body (blue) and 3-body (green) losses. The modes are
    labeled by their eigenfrequencies
    $\omega_\nu = \omega \sqrt{\nu(\nu+1)/2}$ and we only consider
    $\nu \ge 2$.
    Symbols: calculation based on the Legendre polynomials
    of Eq.(\ref{eq.gnuharm1D}), inserted into
    Eqs.~(\ref{eq.calA}, \ref{eq.calB}). 
    Solid lines: large-$\nu$ approximation
    given by Eq.~(\ref{eq.yinftyharmnugrand}) with values
    $y_\infty = 3/4, 45/56, 525/748$ for $j = 1, 2, 3$.}
  \label{fig.yinftyharm}
\end{figure}

We consider a 1D gas confined in a harmonic potential of 
trapping frequency $\omega$. We assume for simplicity a
pure 1D situation with $\mu = g n = m c^2$. In the Thomas-Fermi approximation,
the mean density profile is  
\begin{equation}
n_0(z)=n_p(1-(z/\RTF)^2)
\,, \quad
|z| \le R
\label{eq.noharm}
\end{equation}
where $n_p$ is the peak density and
$\RTF = \sqrt{2gn_p/(m \omega^2)}$ is the axial radius of the quasi\-condensate. 
From Eq.(\ref{eq:wave-eqn-gnu}), we recover the
  known result that  the hydrodynamic modes are described by
  the Legendre 
polynomials $P_\nu$,
and the eigenfrequencies are 
$\omega_\nu=\omega \sqrt{\nu(\nu+1)/2}$~\cite{Ho_1999, Petrov_2000}.
A trivial calculation using {$N_{\rm tot} = \frac{4}{3} n_p R
\propto  c_p^{3}$} and the substitution $z = \RTF \cos\alpha$ 
gives  
${\cal C} = \int_0^{\pi/2}\!d\alpha \, \sin^{2j+1}\alpha
= 2/3, 8/15, 16/35$ for $j = 1, 2, 3$. 
To compute ${\cal A}$ and  ${\cal B}$, one needs the 
  exact expression of $g_\nu$, which according to the normalisation~(\ref{eq:g-nu-normalisation}) can be written
\begin{equation}
g_\nu(z) = 
	\sqrt{\frac{ \hbar \omega_\nu }{ 2 g R } }
	\sqrt{2\nu+1}
	P_\nu(z/R)
	\,.
\label{eq.gnuharm1D}
\end{equation}
Inserting this expression, together with Eq.~(\ref{eq.noharm}), 
into the integrals~(\ref{eq.calA}) and~(\ref{eq.calB}),
we find that
${\cal A}$,  ${\cal B}$, and ${\cal C}$ are time-independent.
Thus $y$ tends at long times towards the  asymptotic value
$y_\infty=j{\cal B}/(j{\cal A}-{\cal C})$.
For large $\nu$, one can use the asymptotic 
expansion~\cite{DLMF}
\begin{equation}
P_{\nu}\left(\cos\alpha\right) 
\simeq
\left(\frac{2}{\pi(\nu + \frac12)\sin\alpha}\right)^{1/2}\!
\cos\phi_{\nu} 
,
\label{eq:Legendre-asymptote}
\end{equation}
with
$
\phi_{\nu} = (\nu + \tfrac{1}{2}) \alpha - \tfrac{1}{4} \pi
$.
Moreover the fast oscillations of $P_\nu(\cos\alpha)$  
can be averaged out in 
the calculation of the coefficients 
${\cal A}$ and ${\cal B}$. Then ${\cal A}$ and ${\cal B}$ no longer
   depend on $\nu$, so that $y_\infty$ is identical for all modes, 
and we find 
\begin{equation}
y_\infty \simeq
\frac{\frac{j}{\pi}\int_0^{\pi/2}\!d\alpha \,\sin^{2j}\alpha}
{\frac{2j}{\pi}\int_0^{\pi/2}\!d\alpha\, \sin^{2j-2}\alpha
- \int_0^{\pi/2}\!d\alpha \,\sin^{2j+1}\alpha}
\label{eq.yinftyharmnugrand}
\end{equation}
For one- and three-body losses, this gives
$y_\infty = 3/4 = 0.75$
and $y_\infty = 525/748 \simeq 0.701$, respectively.
This asymptotic result is compared to calculations using the expression
Eq.~(\ref{eq.gnuharm1D}) in Fig.~\ref{fig.yinftyharm}. 
We find very good agreement as soon as the mode index is
larger than 5.

To conclude this example, we come back to the diffusive 
dynamics of the `phase quadratures' $p_\nu$
we neglected so far.
In the case of one-body losses, however, it happens that
the integral~(\ref{eq.dpnu}) does not converge: while the
mode function $g_\nu(z)$ [Eq.(\ref{eq.gnuharm1D})] remains
finite at the condensate border $z \to \pm R$, the integrand
$n_0^{j-2}(z) g_\nu^2(z)$ is not integrable for $j = 1$.
This is actually an artefact of the
hydrodynamic approximation, which breaks down at the border of the condensate.

\begin{figure}[h!]
  \centerline{
  \includegraphics[width=0.55\columnwidth]{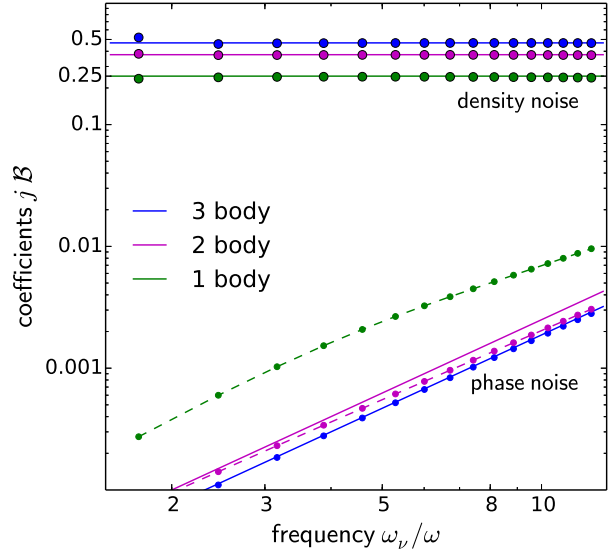} }%{dens-phase-noise_4.eps}}
\caption[]{Diffusion of density and phase quadratures associated
with many-body loss in a one-dimensional gas trapped in a harmonic potential.
We plot the dimensionless 
coefficients $j {\cal B}$ [Eq.(\ref{eq.calB})] 
and $j {\cal B}_\varphi$ [Eq.(\ref{eq.Bphi})] that are proportional
to the shot noise projected onto the corresponding quadratures.
Symbols: numerically computed mode functions, improving upon the
hydrodynamic approximation. Solid lines: approximate results based on the
Legendre modes~(\ref{eq.gnuharm1D}). Dashed lines: guide to the
eye. Parameters: strictly 1D equation of state $\mu = g n$,
peak chemical potential
$\mu_p \approx g n_p = 100\, \hbar\omega$.
}
\label{fig.noise-density-phase}
\end{figure}

We have performed numerical calculations of the
collective excitations by solving the Bogoliubov equations~\footnote{For the condensate
  wave-function, we also went beyond the Thomas-Fermi approximation by allowing
  for a `spill-over' of the
condensate density beyond the inverted parabola.}.
The mode functions $g_\nu( z )$ are defined
according to Eq.(\ref{eq:def-fpm-numerics}):
they extend smoothly beyond the Thomas-Fermi radius and
match well with the Legendre polynomials~(\ref{eq.gnuharm1D})
within the bulk of the gas.
The resulting values for the parameter ${\cal B}$
[Eq.(\ref{eq.calB})] are shown in Fig.\ref{fig.noise-density-phase}:
they depend very weakly on the mode index $\nu$ and are well
described by 
the approximate calculation based on the
Legendre modes mentioned after Eq.(\ref{eq:Legendre-asymptote})
(solid lines). 
In the lower part of the figure, the corresponding values
for the diffusion coefficient originating from phase noise
are shown, namely the parameter 
\begin{equation}
{\cal B}_\varphi = 
\frac{\hbar\omega_\nu }{ 8 m c_p^2 }
  \int\!d^{d}{\bf r} \frac{ n_0^{j-2} }{ n_p^{j-1} } g_\nu^2({\bf r}) 
  \,.
\label{eq.Bphi}
\end{equation}
They remain at least one order of magnitude below.
For losses involving more than one particle,
  the { approximation}, under which the functions $g_\nu$ are given by the Legendre
  polynomials, gives a convergent integral in Eq.(\ref{eq.Bphi}).
  The result
    is shown as solid lines for two- and three-body losses, where we made the additional
    approximation Eq.~(\ref{eq:Legendre-asymptote}) on the Legendre functions
    and we averaged out the oscillating part.
  {We find that the Legendre  approximation} performs better for three-body losses
  {than for 2-body losses, which is expected since}
a stronger weight is given to the bulk rather
than the edge of the condensate.
{In conclusion of this numerical study, we verified the validity of the assumption that,
for phonon modes,
the phase diffusion term gives negligible contribution to the evolution of $y$.
This term becomes noticeable when one leaves the phonon regime $\hbar\omega_\nu \ll mc^2$. Then, one should go beyond the hydrodynamic Hamiltonian~Eq.(\ref{eq.Hydro})
to properly compute the mode dynamics.\footnote{A full treatment going beyond the hydrodynamic
approximation has been performed for one-body losses in homogeneous 1D quasi-condensates~\cite{johnson_long-lived_2017,grisins_degenerate_2016}.}}

\section{Conclusion}
In this paper, we construct a stochastic model
to describe the effect of losses
on the hydrodynamic collective modes of
condensates or quasi\-condensates. Explicit formulas
for cooling and diffusion of the density and phase
quadratures are derived. They provide the behaviour of
the mode temperature $T$ with time.
We show that $T$ becomes proportional to the energy scale
$m c^2$ where $c$ is the hydrodynamic speed of sound.
The asymptotic ratio
$k_B T/(mc^2)$ is computed explicitly in different situations
and for different $j$-body processes.
{These results} are in good agreement with recent experiments
in our group~\cite{schemmer_cooling_2018} where three-body
losses provided the dominant loss channel.

This work raises many different questions and remarks.
First, it is instructive to investigate the evolution of
  the ratio $D = \hbar^2 n^{2/d}/(m k_B T)$, where
  $d$ is the gas dimension, since $D$
  quantifies the quantum degeneracy of the gas.\footnote{Note
  however that the temperature used in the definition of $D$ 
  refers to the phononic modes only.} Let us focus for simplicity 
  on a homogeneous system and use $mc^2 = gn$. 
  Once the ratio $k_B T / (mc^2)$ has become stationary, 
  we find that $D$ increases in time for 3-dimensional gases, 
  while it decreases for one-dimensional gases.
  Starting with a 1D Bose gas in the quasi-condensate regime, 
losses let the quantity $D\gamma$ reach a stationary value of order one,
but increase the dimensionless interaction parameter $\gamma=mg/(\hbar^2 n)$.
When $\gamma$, from values much smaller, approaches 1, the gas 
lies at the crossover between four regimes:  the quasi-condensate ($\gamma\ll 1$, $D \sqrt{\gamma}\gg 1$),
  the quantum-degenerate ideal Bose gas ($D \sqrt{\gamma}\ll 1$, $D \gg 1$),
  the non-degenerate ideal Bose gas ($D {\gamma}^2 \ll 1$, $D \ll 1$) and the
  Tonks-Girardeau
  regime ($\gamma \gg 1$, $D {\gamma}^2 \gg 1$). At later times, one expects the cloud
  to leave the
  quasi-condensate regime and we believe it becomes a non-degenerate
  ideal Bose gas.
Second, the effect of losses on high-frequency modes, not described by our
  hydrodynamic model, leads a priori to higher
temperatures; this was investigated for 1D gases subject to one-body
losses~\cite{johnson_long-lived_2017}. The gas is then described
by a generalised Gibbs ensemble
where different collective modes experience different temperatures.
This non-thermal state is even long lived in 1D quasi\-condensates~\cite{johnson_long-lived_2017}.
While the calculations presented here are formally valid for higher dimensions,
efficient coupling between modes may reduce their relevance, since such
  coupling favours
  a common temperature.
  It is an open question whether our methods
could be extended to the case of evaporative cooling where the one-body loss
rate is energy- or position-dependent. This mechanism may play a role in experiments
where temperatures as low as $k_B T \approx 0.3\,m c^2$ have been observed, lower than the
predicted temperatures for uniform losses~\cite{rauer_cooling_2016}.
Finally, it would be interesting to extend this work to different regimes of the gas.
For instance, one may ask how the effect of losses transforms
as one goes from a quasi-condensate to the ideal gas regime.
The approximation of weak density fluctuations then clearly becomes invalid.
One could also investigate losses at even lower densities, where the 1D gas enters
the fermionised (or Tonks-Girardeau) regime. Here, ones expects that the losses act
in a similar way as in a non-interacting Fermi gas. One-body losses, for example,
should then produce heating, since the temperature increases as the degeneracy
of an ideal Fermi gas decreases.
Finally, it would be interesting to investigate whether the results presented here
may also cover interacting Fermi gases in the superfluid regime.

\section*{Acknowledgements}
M. S. gratefully acknowledges support by the 
\emph{Studienstiftung des deutschen Volkes}. 
% German Academic Scholarship Foundation.
This work was supported by R\'egion \^{I}le de France (DIM NanoK, Atocirc project). The work of C. H. is supported by the
\emph{Deutsche Forschungsgemeinschaft} (grant nos. Schm 1049/7-1 and Fo 703/2-1).
%Acknowledgements should follow immediately after the conclusion.
% TODO: include funding information                                                                                             
%% \paragraph{Funding information}
%% This work was supported by R\'egion
%% \^{I}le de France (DIM NanoK, Atocirc project). Authors are required to provide funding information, including relevant agencies and grant
%% numbers with linked author's initials. Correctly-provided data will be linked to funders
%% listed in the
%% \href{https://www.crossref.org/services/funder-registry/}{ Fundref registry}.

\begin{appendix}

\section{\protect Reduction to low-dimensional hydrodynamics}
% Derivation of the hydrodynamic Hamiltonian}
\label{a:lowD-hdyn}

As mentioned in the main text, we assume the loss process is slow enough so that, first, the
mean profile at each time is very close to the equilibrium profile with the same atom number,
and second, we can safely neglect any mean velocity field when computing the time evolution
of the fluctuating fields $\delta n$, $\varphi$.
The evolution equations $\partial\delta n/\partial t$ and  $\partial \varphi/\partial t$
are thus, at a given time, equal to those for a time-independent quasi-condensate. 
In the purely 3D, 2D and 1D cases, for contact interactions,
  we can use the well known results based on Bogoliubov theory. We then find that
  the equation of state
     takes the form $\mu=gn$ and 
  $\partial\delta n/\partial t$ and  $\partial \varphi/\partial t$
  derive from Eq.~(\ref{eq.Hydro}) for the long-wavelength modes. 

Let us now consider the case where the gas is confined
strongly enough in
1 or 2 dimensions, such that the relevant low-lying excitations are
of planar or axial nature. We allow, however, 
for a transverse broadening of the wave function under the effect of
interactions. We show below that the equations of motion for the
slow phononic modes, for which the transverse shape adiabatically follows the 
density oscillations, also derive from Eq.~(\ref{eq.Hydro}). The proof given here is complementary
to Refs.\cite{Stringari_1998, salasnich_effective_2002} because
it does not need an explicit model about the shape of the transverse
wave function.
In order to simplify the
notations, we restrict ourselves to  the quasi-1D situation. The derivation can be
easily translated to quasi-2D situations.

We thus consider a gas confined in a separable potential consisting of a
 strong transverse confinement and a smooth longitudinal confinement.
The equilibrium density distribution of the quasi-condensate is $|\phi_0(x,y,z)|^2$ where
the real function $\phi_0(x,y,z)$ obeys the stationary Gross-Pitaevskii equation
\begin{equation}
  \left(-\frac{\hbar^2}{2m}\partial_z^2 -\frac{\hbar^2}{2m}\Delta_\perp + V_\perp(x,y)+V(z) + g |\phi_0|^2 -\mu_p \right)\phi_0=0.
\label{eq:GPE}
\end{equation}
Here $g=4\pi\hbar^2a/m$ is the 3D coupling constant with $a$ the zero-energy scattering length.
Within the Bogoliubov theory, the evolution of excitations is governed by the
equations~\cite{PitaevskiiStringari}
\begin{equation}
  \left \{
    \begin{array}{rcl}
      i\hbar\partial_t \tilde{f}^+&=& 
      \displaystyle
      \left(-\frac{\hbar^2}{2m}\partial_z^2 -\frac{\hbar^2}{2m}\Delta_\perp + V_\perp(x,y)+V(z) + g |\phi_0|^2 -\mu_p \right) \tilde{f}^- 
      \\[0.5ex]
      %\label{eq.evolf+f-f+}\\
      i\hbar \partial_t \tilde{f}^-&=& 
      \displaystyle
      \left(-\frac{\hbar^2}{2m}\partial_z^2 -\frac{\hbar^2}{2m}\Delta_\perp + V_\perp(x,y)+V(z) + 3g |\phi_0|^2 -\mu_p \right) \tilde{f}^+
      %\label{eq.evolf+f-f-}\\
      \end{array}
    \right .
    \label{eq.evolf+f-}
    \end{equation}
The field operators are half sum and difference of the fluctuating 
field operators $\delta\psi$ and $\delta\psi^\dag$. $\tilde{f}^+$ is linked to density fluctuations 
 and $\tilde{f}^-$ to phase fluctuations. 
%% The mode functions $\tilde{f}^\pm = u \pm v$
%% are %proportional to
%% the sum and difference of the usual Bogoliubov
%% functions~\footnote{We choose the normalisation  $\int\!d^3r (u^2 - v^2) = 1$.}:
%% $\tilde{f}^+$ is linked to density fluctuations 
%% and $\tilde{f}^-$ to phase fluctuations. w

Since we assume
that the axial variation is slow compared to the transverse one, the solution $\phi_0$ 
can be approximated by a function $\psi$ that  
depends on the axial coordinate $z$ only via a local chemical
potential
\begin{equation}
\phi_0( x, y, z ) \simeq \psi(x, y; \mu)
\,,\qquad
\mu = \mu_p - V(z)
\label{eq:phi0-separation}
\end{equation}
Here, $\psi$ solves the Gross-Pitaevskii
equation for an axially homogeneous system:
\begin{equation}
  \left(-\frac{\hbar^2}{2m}\Delta_\perp + V_\perp(x,y) 
  + g |\psi|^2 - \mu \right)\psi = 0.
  \label{eq.psimu}
\end{equation}
This procedure is consistent, e.g., with making the Thomas-Fermi
approximation in the axial direction. 
Solving this equation yields the local chemical potential
as a function of the axial (average) density $\mu = \mu( n_0 )$ 
with
\begin{equation}
n_0( z ) = 
	\int\!dx dy\, |\phi_0(x, y, z)|^2
\simeq
	\int\!dx dy\, |\psi(x,y; \mu)|^2
\label{eq:def-densite-1D}
\end{equation}
This motivates the following separation \emph{Ansatz} for the
Bogoliubov functions in Eq.(\ref{eq.evolf+f-}):
\newcommand{\fPlus}{F^+}
\newcommand{\fMinus}{F^-}
\begin{equation}
  \left \{
  \begin{array}{l}
    \tilde{f}^+ = \partial_\mu \psi \partial_n \mu \fPlus
    \\[0.5ex]
    \tilde{f}^- = i\phi_0\fMinus
  \end{array}
  \right .
  \label{eq.ansatzf+f-}
\end{equation}
where the functions $\fPlus$ and $\fMinus$ depend only on $z$
and the derivative $\partial_n \mu$ is evaluated at the local density 
$n_0$.
Inserting this into the second line of Eq.(\ref{eq.evolf+f-}), we find
\begin{equation}
  -\phi_0\hbar\partial_t\fMinus
  =
  \left(-\frac{\hbar^2}{2m}\partial_z^2 -\frac{\hbar^2}{2m}\Delta_\perp + V_\perp(x,y)+V(z) + 3g |\phi_0|^2 -\mu_p \right)
  \left( \partial_\mu \psi \partial_n \mu \fPlus\right)
  \label{eq.partialF-}
\end{equation}
The action of this operator on $\partial_\mu \psi$ can be
worked out by differentiating Eq.~(\ref{eq.psimu}) versus $\mu$:
this gives
\begin{equation}
\left( -\frac{\hbar^2}{2m}\Delta_\perp + V_\perp(x,y)+ 3g |\psi|^2 -\mu \right)\partial_\mu \psi = \psi \simeq \phi_0
\end{equation}
Eq.(\ref{eq.partialF-}) thus simplifies into
\begin{equation}
  -\phi_0\hbar\partial_t\fMinus
  =
  - \frac{ \hbar^2 }{ 2 m } \partial_z^2\left( \partial_\mu \psi  \partial_n \mu \fPlus\right)
  + \phi_0\partial_n \mu \fPlus
\label{eq:partialF-etape-2}
\end{equation}
To find a closed equation for the axial dynamics, we multiply
with $\psi(x, y; \mu )$ and integrate over the transverse coordinates.
Using Eq.(\ref{eq:def-densite-1D}) and its derivatives with
respect to $\mu$ and $z$, we find the identities
\begin{equation}
\int\!dxdy\, \phi_0 \partial_\mu \psi = \frac12 \partial_\mu n_0 
	= \frac{ 1 }{ 2 \partial_n \mu }
\,,
\qquad
\int\!dxdy\, \phi_0 \partial_z \phi_0 = \frac12 \partial_z n_0
\,.
\label{eq:astuces-projection}
\end{equation}
Using the first one, Eq.(\ref{eq:partialF-etape-2}) becomes:
\begin{equation}
  - \hbar\partial_t\fMinus
  = - \frac{ \hbar^2 }{ 4 m n_0 } \partial_z^2 \fPlus
  + \partial_n\mu \fPlus
  \simeq
  -\partial_n\mu \fPlus
  \label{eq.F-ronde}
\end{equation}
where in the second step, we took the long-wavelength limit.

Let us now insert the \emph{Ansatz}~(\ref{eq.ansatzf+f-}) 
into the first line of Eq.(\ref{eq.evolf+f-}):
\begin{equation}
   \hbar\partial_\mu\psi \partial_n\mu \partial_t\fPlus
   =
   \left(-\frac{\hbar^2}{2m}\partial_z^2 -\frac{\hbar^2}{2m}\Delta_\perp + V_\perp(x,y)+V(z) + g |\phi_0|^2 -\mu_p \right)
   \left( \phi_0\fMinus \right)
\end{equation}
The action of the operator in parentheses on $\phi_0$ simply vanishes
because this is
the Gross-Pitaevskii equation~(\ref{eq:GPE}). Since
$\fMinus$ does only depend on the axial coordinate, we are left 
with:
\begin{equation}
  \partial_\mu\psi\partial_n\mu \partial_t \fPlus
  =
  -\frac{\hbar}{m}
  \left(\partial_z\phi_0\right)\partial_z \fMinus
  - \frac{\hbar}{2m}\phi_0\partial_z^2 \fMinus
\end{equation}
We again project out the transverse coordinates and use
the identities~(\ref{eq:astuces-projection}). 
Combining the axial derivatives, we then have
\begin{equation}
  \partial_t \fPlus
  =
  -\frac{\hbar}{m}\partial_z  ( n_0 \partial_z \fMinus )
  \label{eq.F+ronde}
\end{equation}
These calculations illustrate that the \emph{Ansatz} of
Eq.(\ref{eq.ansatzf+f-}) captures well the axial and transverse
dependence of the collective excitations in the low-dimensional gas.
Note in particular how the density fluctuations ($\tilde{f}^+$) are
accompanied 
by density-dependent changes in the transverse wave function.

  To make contact with the hydrodynamic Hamiltonian~(\ref{eq.Hydro}),
  we need to relate $F^+$ and $F^-$ to the low-dimensional
  density and phase fields, $\delta n$ and $\varphi$.
  Bogoliubov theory tells us that three-dimensional density fluctuations are linked to $\tilde{f}^+$ via
  $\delta \rho = 2\phi_0 \tilde{f}^+$.
  Integrating $\delta \rho$ 
  over the transverse plane, replacing $\tilde{f}^+$ by its Ansatz~(\ref{eq.ansatzf+f-})
  and using Eq.~(\ref{eq:astuces-projection}),
  we obtain
  \begin{equation}
    F^+ = \delta n =n-n_0.
  \end{equation}
  Phase fluctuations on the other
  hand are linked to $\tilde{f}^-$ according to $\tilde{f}^-=i\phi_0\varphi$. [Recall
    that the ansatz~(\ref{eq.ansatzf+f-}) assumes a uniform phase in the $x,y$ plane.]
  Comparison with
  Eq.~(\ref{eq.ansatzf+f-})  gives immediately
  \begin{equation}
    F^- = \varphi.
  \end{equation}
  Then Eq.(\ref{eq.F-ronde}) and Eq.(\ref{eq.F+ronde}) are
  precisely the evolution equations derived from the Hamiltonian~(\ref{eq.Hydro}).

\section{\protect Hydrodynamic Bogoliubov modes}
\label{SM}

{Here we consider low-energy modes of
  either a three-dimensional gas or low-dimensional gas,
  whose dynamics is described by the hydrodynamic 
  approximation.}
{More precisely,} we diagonalize the 
Hamiltonian~(\ref{eq.Hydro}), for a
given, time-independent, equilibrium  profile $n_0({\bf r})$.
From Eq.(\ref{eq.Hydro}) we derive the evolution equations
\begin{equation}
  \frac{\partial}{\partial t}
  \left(
  \begin{array}{c}
    \delta n/\sqrt{n_0}\\
    \sqrt{n_0}\varphi
    \end{array}
  \right)
  =
  {\cal L}
  \left(
  \begin{array}{c}
    \delta n/\sqrt{n_0}\\
    \sqrt{n_0}\varphi
    \end{array}
  \right)
  \label{eq.evolhydro}
\end{equation}
where 
\begin{equation}
  {\cal L}=
  \left(
  \begin{array}{cc}
    0 & - \frac{\hbar}{m\sqrt{n_0}}\nabla \cdot \left( n_0 
    \nabla \left( \frac{1}{\sqrt{n_0}} \cdot \right)\right)\\
    - m c^2 / \hbar & 0
    \end{array}
  \right)
\end{equation}
The factors $\sqrt{n_0}$ are convenient to give the two components
the same dimension and to symmetrize the differential operator
that appears in ${\cal L}$.
The two equations derived from Eq.(\ref{eq.evolhydro})
correspond to the hydrodynamic equations provided we identify
$\hbar \nabla \varphi/m$ with the velocity:
the first one is the continuity equation, the second one gives
the Euler equation. 

We build the mode expansion on pairs of 
real functions that form right eigenvectors of ${\cal L}$:
\begin{equation}
{\cal L} \left( \begin{array}{c}
    f^+_\nu \\ i f^-_\nu
    \end{array} \right)
    =
    i \omega_\nu
    \left( \begin{array}{c}
    f^+_\nu \\ i f^-_\nu
    \end{array} \right)
\label{eq:eigenmode-problem}
\end{equation}
Due to symmetry properties of ${\cal L}$, Eq.(\ref{eq:eigenmode-problem}) 
entails the following properties:
({\it a}) 
$(f_\nu^+,-if_\nu^-)$ is a
right eigenvector of ${\cal L}$ of eigenvalue $-i\omega_\nu$;
({\it b}) 
$(if_\nu^-,f_\nu^+)$ is a left eigenvector of same eigenvalue;
and ({\it c}) different right eigenvectors of ${\cal L}$ verify
$\int\!d^{d}{\bf r}\, f_{\nu}^-f_{\nu'}^+=0$.
It is convenient to consider those eigenvectors of ${\cal L}$
which are normalized according to $\int\!d^{d}{\bf r}\, f_{\nu}^-f_{\nu}^+=1$.
This yields the expansions
\begin{equation}
  \left(
\begin{array}{c}
    \delta n/\sqrt{n_0}\\
    \sqrt{n_0}\varphi
    \end{array}
  \right)
  = \frac{ 1 }{ \sqrt{2} }
  \sum_\nu \left \{
  a_\nu \left(
\begin{array}{c}
   f_\nu^+\\
    -i f_\nu^-
    \end{array}
\right)
+ a_\nu^+ \left(
\begin{array}{c}
   f_\nu^+\\
    if_\nu^-
    \end{array}
\right)
\right \}
\label{eq.deltantheta}
\end{equation}
which invert into
\begin{equation}
  a_\nu = \frac{ 1 }{ \sqrt{2} }
  \int\!d^d{\bf r} \left( 
  \frac{ \delta n({\bf r}) }{\sqrt{n_0}} f_\nu^-({\bf r}) + i\sqrt{n_0}\,\varphi({\bf r}) f_\nu^+({\bf r})
  \right)
  \,.
  \label{eq.anu}
  \end{equation}
The normalisation of the eigenvectors and the relation $[\delta n(z),\varphi(z')] = i\delta(z-z')$
ensure 
$[a_{\nu'},a_\nu^\dag] = \delta_{\nu',\nu}$.

We introduce the
function
\begin{equation}
    g_\nu = \sqrt{n_0} \,f_\nu^+
    ,
    \label{eq-B:def-gnu}
\end{equation}
and use the relation
$f_\nu^- =  m c^2 f_\nu^+/(\hbar\omega_\nu)$ that follows
from the eigenvalue problem~(\ref{eq:eigenmode-problem}).
Then the normalisation of $g_\nu$ [Eq.(\ref{eq:g-nu-normalisation})]
follows from that of  $(f_\nu^+,if_\nu^-)$.
Defining the quadratures $x_\nu = (a_\nu+a_\nu^\dag)/\sqrt{2}$ and $p_\nu = - i(a_\nu-a_\nu^\dag)/\sqrt{2}$,
the expansions~(\ref{eq.deltantheta}) give Eqs.(\ref{deltanvsxnu}) 
of the main text.

\section{\protect Numerical calculation}

For the numerical results shown in Fig.\ref{fig.noise-density-phase},
we have solved
the Gross-Pitaevskii equation in a 1D harmonic
trap by minimising the corresponding energy functional: 
this gives a smooth density profile $n_0( z )$. 
The Bogoliubov equations are solved with a finite-difference scheme 
on a non-uniform grid. 
We get a frequency spectrum that coincides to better
than one percent with the Legendre spectrum for all modes
with $\hbar \omega_\nu \lesssim 0.1\,g n_p$ 
($n_p$ is the peak density).
The traditional Bogoliubov modes $u_\nu$ and $v_\nu$ are 
related to the eigenfunctions of Eq.(\ref{eq:eigenmode-problem}) 
by 
\begin{eqnarray}
f^+_\nu &=& \sqrt{2}\,( u_\nu + v_\nu )
\\
f^-_\nu &=& (u_\nu - v_\nu )/\sqrt{2}
\label{eq:def-fpm-numerics}
\end{eqnarray}
Inserting this into Eq.(\ref{eq-B:def-gnu}) gives the 
modes $g_\nu$.
We have checked for phonon excitations with frequencies 
$\hbar\omega_\nu \ll g n_p$,
that the proportionality between $f^+$ and $f^-$ 
[see after Eq.(\ref{eq-B:def-gnu})]
is an excellent approximation in the bulk of the condensate.

\end{appendix}

% \bibliography{losses-j-body,pertes3corps,papierexp}

%\nolinenumbers

\end{document}